\documentclass{aa}
\usepackage{graphics}
\begin{document}

\title{SS433: on the uniqueness of cool relativistic jets}
\author{A.A. Panferov}
\institute{Nicolaus Copernicus Astronomical Center, Bartycka 18, Warsaw,
PL-00716, Poland; panferov@camk.edu.pl\\
Special Astrophysical Observatory, Nizhnij
Arkhyz, 357147, Russia; panf@sao.ru}
\date{Received/Accepted}
\thesaurus{08(08.09.2 SS 433; 09.10.1)}
\maketitle
\begin{abstract}

The relativistic jets of
SS433 are outstanding for their optical thermal radiation.
The radiation is produced by small clouds ($10^8$ cm) whose
lifetime is about $10^3$ times larger than the
gas-dynamical crushing time.
We show that the clouds reside in thermal and dynamical balance
as long as they collisionally interact with the wind of the supercritical
accretion disk. The interaction is caused by the precessional
movement of the jets and takes place only in the sweep-out zone.
Beyond the sweep-out zone the interaction ceases and optical
jets just terminate.
The cloud magnetic field amplified in course of movement 
through a medium could play a role in containing a cloud.
Thus, the clue to the uniqueness of the optical jets 
of SS433 is thought to be their precessional movement, 
which provides an opportunity for collisional interaction of
the clouds with the wind.
\keywords{Stars: individual: SS\,433 -- ISM: jets and outflows}
\end{abstract}

\section{Introduction}
Since the discovery of jets in SS433 in 1978, there has been 
discussion about their uniqueness. By now the zoo of various 
jets has enlarged.
There have already been detected hundreds relativistic jets of
AGNs, low velocity jets of young stars and a dozen
jets of compact stars. Jets of SS433 stand out among the others for
they harbour cool clouds within the relativistic flow. 
It is not clear, however, how the clouds could survive in 
the relativistic jets and be observed. 
It is also worth mentioning, that the thermal fraction is likely
to be dominant in the jets of SS433. 
On the contrary, the lack of observational evidence of thermal gas
in AGN jets strongly restricts its volume-filling factor
($<10^{-8}$) and implies that it cannot be important in the 
energy budget of these jets (\cite{Celotti et al. 1998}). 
These peculiarities make the jets of SS433 unique.

Jets of SS433 are highly collimated, $\theta_\mathrm{j} \le 1^\circ$.4,
mildly relativistic, $v_\mathrm{j}=7.8\cdot 10^{9}$ cm\,s$^{-1}$,
and powerful, with kinetic luminosity
$L_\mathrm{k} \approx 10^{39}$ erg\,s$^{-1}$.
They appear from funnels of the supercritical
accretion disk and follow its precession rotation with a half opening
angle of about $20^\circ$ and a period of $P_\mathrm{pr}=162^\mathrm{d}.5$
(see Fig.~\ref{wall}). At the distance of a few 
$10^{12}$ cm the jets rapidly cool and radiate in the X-ray domain. 
The X-ray emission does not show any structure of the jets.
Therefore, the X-ray jets are thought to be continuous 
(\cite{Stewart et al. 1987}).
By contrast, the optical jets are observed as fragmented (\cite{Vermeulen
 et al. 1993b}). Their clumpness may be roughly deduced from energetic
considerations. The gas must be dense enough to radiate the bulk of
the emission:
\begin{equation}
\label{lha}
L_\mathrm{H_\alpha}=\epsilon_\mathrm{H_\alpha}n_\mathrm{e}^2\,f\,V,\nonumber
\end{equation}
and mass loading of the jet by dense gas must approximately satisfy:
\begin{equation}
\label{lki}
L_\mathrm{k}=\dot{M}_\mathrm{j} v_\mathrm{j}^2,\nonumber
\end{equation}
where $V \approx \theta_\mathrm{j}^2 R^3/4$ is the volume of the jet,
$R$ is its length, $n_\mathrm{e}$ is electron
density, $\epsilon_\mathrm{H_\alpha}$ is the $\mathrm{H_\alpha}$
emission rate. Thus, one finds that 
$n_\mathrm{e}\approx 5\cdot 10^{11}$ cm$^{-3}$ and a
volume-filling factor of $f\approx
10^{-5}$ for appropriate parameters: luminosity in $\mathrm{H_\alpha}$ line
$L_\mathrm{H_\alpha}=10^{36}$ erg\,s$^{-1}$, total jet kinetic luminosity
$L_\mathrm{k}=10^{39}$ erg\,s$^{-1}$, $\epsilon_\mathrm{H_\alpha}=
10^{-24}$ erg\,cm$^3$\,s$^{-1}$,
$R=R_\mathrm{e} \approx 8.4\cdot 10^{14}$ cm --- length of {\em e}-folding
$\mathrm{H_\alpha}$ line brightness.
It seems that the jets become lumpy in
the gap between the X-ray part and the optical jet beginning at about
$R_\mathrm{in}=2.3\cdot 10^{14}$ cm (\cite{Borisov Fabrika 1987}).
The gas could
condense there due to a thermal instability as Brinkmann et al. 
(\cite{Brinkmann et al. 1988}) have shown in a simulation of the 
thermal evolution of the X-ray jets. The instability
takes place only in the interior jet, until a distance of $10^{13}$ cm. 
The parameters of the clouds found in their work are close to
those obtained by Panferov \& Fabrika (\cite{Panferov Fabrika 1997})
from the Balmer decrements of the jets:
hydrogen density $n \ge 10^{13}$ cm$^{-3}$ and size of the clouds
$l\le 10^8$ cm. It should be noted that these parameters as well as
the filling factor $f \approx 4\cdot 10^{-6}$  found there are more
appropriate for the distance of maximum optical brightness of the jets
$R_\mathrm{m} \approx 4\cdot 10^{14}$ cm.
These parameters will be representative
ones in what follows. The clouds are grouped
in clusters with sizes of about $10^{12}$ cm (\cite{Panferov Fabrika 1997}).
Still bigger structures are quite possible in the jets
as may be deduced from the structure of the jet emission line.

Brown et al. (\cite{Brown et al. 1995})
consider cloudy structure of the jets as a result of a temporal
switch on and off of conditions for thermal instability. They
found that the parameters of SS\,433 are just suitable for
the instability to occur. However, the instability can work with a wide
range of the parameters and in SS\,433 jets the conditions for 
cloud generation exist rather continuously in time.
Indeed, the data on optical emission of the
jet (e.g. \cite{Vermeulen et al. 1993b}) demonstrate that the generation
of the clouds in the jets is continuous and the disappearance
of the optical jets happens very rarely. The intermittent 
structure of the jets is natural if the radiating clouds form
in such {\em in situ} variable process as the thermal instability.
The optical jets terminate
at a distance of about $R_\mathrm{j}=3\cdot 10^{15}$ cm (\cite{Borisov
Fabrika 1987}) and then are observed to be quasi-continuous in radio
(\cite{Vermeulen et al. 1993a}). It is observed as separated emission
peaks grow and decay at a unchanged wavelength reflecting
an evolution of ``bullet'' along the jets (\cite{Borisov Fabrika 1987};
\cite{Vermeulen et al. 1993b}). Thus, the clouds form in the interior
jets and further exist a long  time  giving a signature of
the optical jets.

The actual problem is how do the clouds evolve and survive in the
relativistic jets? As long as the clouds are involved in a motion
through a medium, that is, the dense wind of the accretion
disk, the applied ram pressure leads to
gas-dynamical instabilities that try to disrupt
the clouds on a crushing time scale (e.g., \cite{Jones et al. 1996}):
\begin{equation}
\label{tcr}
t_\mathrm{cr}=\frac{l}{v_\mathrm{j}}\sqrt{\frac{\rho}{\rho_\mathrm{w}}},
\end{equation}
where $\rho$ is a cloud density,  $\rho_\mathrm{w}$ is a 
wind density. For the  clouds of 
the jets of SS\,433 this time is on the order of $10^2$ s. 
Besides this, the clouds undergo  an evaporation in the hotter
intercloud medium. The evaporation characteristic time may
be rather short, too (see Eq.(\ref{tev})).
Namely the capability of the clouds to survive
permits us opportunity to observe the jets of SS\,433 in the optical.
In this paper we show that the collisional interaction that seems to be
the destructive one is a most necessary condition of the clouds existence. 

In Sect.2 we show that the collisional interaction may be
comparable or even more essential than radiation for the
thermal state of the clouds. Although this
has been concluded by another authors (e.g., \cite{Brown et al. 1995}),
we use here the last data on the jets  parameters (\cite{Panferov Fabrika
1997}) and on parameters of UV radiation of the system --- a possible source
of heat of the clouds (\cite{Dolan et al. 1997}). In Sect.3 we discuss
evolution and confinement of the clouds.

\section{Thermal balance of clouds}
There are two main ways to heat the clouds: by interaction of jets with
the gaseous wind of the supercritical accretion disk and by radiation
of the system (\cite{Davidson McCray 1980}).
Here we consider competition of these two mechanisms. As long as the
cloud sound-crossing
time is shorter than the dynamical time-scale $t_\mathrm{d}=r/
v_\mathrm{j}$ of the jet,
the clouds are in pressure equilibrium with surroundings. The
influence of ram pressure of the wind on the equilibrium is most essential
(see Sect.3). Density in the wind scales as $\propto r^{-2}$, where
$r$ is the radial distance from the source of the jets.
Under plausible conditions of approximate constancy
of temperature and mass of a cloud this implies the 
following dynamical dependences of cloud parameters:
\begin{eqnarray}
\label{dyn}
n=1.6\cdot 10^{14}\, r_{14}^{-2}\ {\rm cm}^{-3}, \nonumber \\ 
l=4\cdot 10^7\, r_{14}^{2/3}\ {\rm cm}, \hspace{.72cm} \nonumber \\ 
f=1.6\cdot 10^{-5}\, r_{14}^{-1},\hspace{.91cm}
\end{eqnarray}
where $r_{14}=r/(10^{14}\ {\rm cm})$ and
we have used the values of the cloud parameters at $R_\mathrm{m}$,
given in Sect.1.
In what follows, we assume these to be valid everywhere along 
the optical jets.

Let us compare the rates of cloud heating by the impinging gas 
and by radiation.
The jets sweep out the gas of the wind due to the change of the 
jet direction in the course of the precession and nodding
motions and a small amplitude jitters of unclear nature
(\cite{Margon Anderson 1989}).
The energy deposition into a cloud by crossing protons
with energy $\epsilon_\mathrm{p}=m_\mathrm{p} v_\mathrm{j}^2/2$
depends on the collisional thickness, measured by the ratio
$\sigma_\mathrm{c} = N/N_\mathrm{s}$, where $N = n\,l = 6.4\cdot 10^{21}
r_{14}^{-4/3}$ cm$^{-2}$
is the column density of the cloud, and $N_\mathrm{s}$ is 
the column density required to stop a proton.
The rate of energy loss of a high energy proton in an ionized gas
is approximately (\cite{Ginzburg Syrovatskii 1964}):
\begin{equation}
\label{gam}
\Gamma_\mathrm{p}=\frac{2.2\cdot 10^{-8}\, K_\mathrm{c} n}
{v_\mathrm{j}}\left(1+1.21 \frac{m_\mathrm{p}}{m_\mathrm{e}}
\frac{k\, T}{\epsilon_\mathrm{p}}\right)^{-3/2}\ {\rm erg\, s^{-1}},
\end{equation}
for temperature $k\,T \ll m_\mathrm{e} c^2$, $k\,T \ll
\epsilon_\mathrm{p}$. The factor
$K_\mathrm{c}=30\,\Lambda/n_\mathrm{e} n \sim 1$, where $\Lambda$ is
the Coulomb logarithm. In an ionized plasma almost all of the lost energy
goes into heating the  electron gas.
The stopping column density of a proton with energy $\epsilon_\mathrm{p} =
32$ MeV is:
\begin{equation}
\label{ns}
N_\mathrm{s}=n\, v_\mathrm{j} \frac{\epsilon_\mathrm{p}}
{\Gamma_\mathrm{p}} \approx 1.4\cdot 10^{23}\
{\rm cm^{-3}}.
\end{equation}
When $N < N_\mathrm{s}$, which is satisfied for the entire jet
(see Eq.(\ref{colden})), the cloud heating rate by the fast protons is:
\begin{equation}
\label{hc}
H_\mathrm{c} \approx \sigma_\mathrm{c}\,\epsilon_\mathrm{p} n_\mathrm{w}
v_\mathrm{j} l^2 =\epsilon_\mathrm{p} n_\mathrm{w} v_\mathrm{j} n\,l^3/
N_\mathrm{s}.
\end{equation}
This is lowest estimate. The magnetic field may be amplified to an
equipartition value at the surface of a cloud moving through a medium
with a weak magnetic field (\cite{Jones et al. 1996}). For assumed
in Eq.(\ref{mf}) magnetic field the gyro-radius of a 32 MeV
proton is $3\cdot 10^5\, r_{14}$ cm.
This is smaller than the clouds radius, therefore
the trajectories of protons may be tangled and the heat input boosted.

The clouds could also be heated by the collimated radiation
from funnels of the accretion disk
(\cite{Arav Begelman 1993}; \cite{Panferov Fabrika 1993})
and by the disk UV radiation. The heating is realized by
$\delta$-electrons --- high energy electrons broken away
from an atom due to ionization. Already for ionization
degree $\ge 0.5$, almost all the energy of $\delta$-electrons
is utilized in the heating of the electron gas of a cloud.
We take into account radiation
only with frequency beyond the Lyman edge. A cloud intercepts radiation
flux with density $j_\nu = \pi (R_\star/r)^2 B_\nu$, where
$R_\star$ is radius of the radiation source, $B_\nu$ is the Planck
function, and relativistic factor $\gamma^{-3}=(1-
v_\mathrm{j}^2/c^2)^{3/2} \sim 1$ is omitted.
The part of the radiation which is absorbed is $1-e^{-\tau}$, where the
optical thickness is $\tau = \sigma\, n_\mathrm{H}\, l
=\sigma\, (1-\zeta)\, n\, l$, the ionized fraction is
$\zeta = n_\mathrm{p}/(n_\mathrm{p}+n_\mathrm{H})$.
Then the radiative heating rate
of one cloud is:
\begin{equation}
\label{hr}
H_\mathrm{r}=l^2\int_{\nu_1} \mathrm{d}\nu\,j_\nu (1-e^{-\tau_\nu}),
\end{equation}
where $\nu_1$ is frequency of Lyman edge. We adopt
here $\sigma_\nu \approx 1.73\cdot 10^{-22}\epsilon_\mathrm{keV}^{-8/3}$
cm$^2$, for photon with energy $\epsilon=\mathrm{1keV}\cdot
\epsilon_\mathrm{keV}$ in the range 
13.6 eV -- 12.4 keV, for cosmic abundances (\cite{Cruddace et al. 1974}).

Parameters of the collimated radiation are not well known since 
its orientation to the Earth is unfavourable. Here we adopt rather rough
estimates for its luminosity $L_\mathrm{X}\sim 10^{39}$ erg\,s$^{-1}$
and temperature
$T_\mathrm{X} \sim 1$ keV (\cite{Arav Begelman 1993}; \cite{Panferov
Fabrika 1993}).
Then the radius of a source of the collimated radiation, in blackbody
approach, is $R_\mathrm{X}\approx 2\cdot 10^7$ cm. The supercritical
accretion disk is a source of near isotropic UV
radiation, its parameters in blackbody approach are:
$T_\mathrm{UV}\approx 72000$ K,
$R_\mathrm{d} \approx 1.5\cdot 10^{12}$ cm (\cite{Dolan et al. 1997}).
For given parameters and $\tau < 1$ Eq.(\ref{hr}) gives
$H_\mathrm{r}^\mathrm{UV}/H_\mathrm{r}^\mathrm{X}
\approx 3\cdot 10^5$, i.e.
heating by UV radiation of the disk could dominate heating
by the X-ray collimated radiation. This ratio might be bigger because
the clouds have rather $\tau > 1$ for the UV radiation,
However situation is complicated by the screening
of the clouds in a jet.

The cross-over path length of some
particle moving in the direction of the jet velocity
 across the jet is determined by the curvature of the jet.
 The motion of the jet caused by nodding is more rapid than
that caused by precession and its rate is 
$\dot \phi \approx 5.7\cdot 10^{-7}$ rad\,s$^{-1}$
(\cite{Borisov Fabrika 1987}).
Using this rate we obtain the cross-over length for a jet with fixed
pattern
$l_\mathrm{f}=v_\mathrm{j}\theta_\mathrm{j}/\dot \phi = 3.4\cdot 10^{14}$
cm, with accuracy of factor 2.  Column density along this path is:
\begin{eqnarray}
\label{colden}
N_\mathrm{f}=\int_r^{r+l_\mathrm{f}} \mathrm{d}r\, f\,n \approx
\frac{1.3\cdot 10^{23}}{r_{14}^2} \hspace{3cm} \nonumber \\
\hspace{2cm} \times \left(1- 
\frac{r_{14}^2}{(r_{14}+l_{\mathrm{f}\,14})^2}
\right)\,  {\rm cm}^{-2}.
\end{eqnarray}
Consequently, we have
$N_\mathrm{f} < N_\mathrm{s}$ everywhere in the optical
jets and the jets are
transparent for the rapid protons. So, the screening effect
is not important for the collisional heating. Meanwhile the
screening factor of clouds in the jet $N_\mathrm{f}/N$ may be
essential for the UV radiation. We now determine $N_1=1/\sigma_\nu
(1-\zeta)$ as the
column density of a cloud layer where
$\tau=1$ for the UV radiation. Then the screening factor
for the UV radiation is $N_\mathrm{f}/N_1 \sim 10^6 (1-\zeta)$
and essentially depends on the ionization degree. Therefore 
we will consider heating $H_\mathrm{r}^{UV}$ of the cloud by UV radiation
as an upper limit of radiative heating (\ref{hr}). Really,
this heating is bigger than
mean heating rate of one cloud, on average over the whole jet,
in $N_\mathrm{f}/N_1$ times;
the opacity of the jet to UV radiation
causes a dominance of the collimated radiation in the heating of
screened clouds.

The ratio of the heating rate from fast protons (Eq.(\ref{hc}))
to the radiation heating rate (Eq.(\ref{hr})) is:
\begin{equation}
\label{rat}
\frac{H_\mathrm{c}}{H_\mathrm{r}^\mathrm{UV}}=
\frac{\epsilon_\mathrm{p} v_\mathrm{j} n\,l\, n_\mathrm{w}
r^2}{\pi\,N_\mathrm{s} R_\star^2
\int_{\nu_1} \mathrm{d}\nu B_\nu^\mathrm{UV}} \approx 10\, r_{14}^{-4/3},
\end{equation}
where we have used the reduced form of Eq.(\ref{hr}) for $\tau>1$,
substituting 1 instead of $1-e^{-\tau_\nu}$,
and used density of wind of the accretion disk:
\begin{equation}
\label{niso}
n_\mathrm{w}=\frac{\dot M_\mathrm{w}}{4\,\pi\,m_\mathrm{p}
\,r^2 v_\mathrm{w}} \approx 1.5\cdot 10^8\, r_{14}^{-2}\ {\rm cm}^{-3}.
\end{equation}
The mass loss rate and wind velocity are taken to be 
$\dot M_\mathrm{w}=10^{-4}\, M_\odot/$y (\cite{van den Heuvel 1981})
and $v_\mathrm{w}=2000$ km/s
(from the width of the emission lines) respectively.
Thus the heating of the jet by fast protons is comparable at least
with the radiation heating. Their rates become equal at a distance
$5.6\cdot 10^{14}$ cm, until which the main bulk of the heat input
into the jet takes place and equals to 64\% of the whole input
given by Eq.(\ref{hj}).
The collisional heating may be rather much bigger than radiative one,
because the value of $H_\mathrm{r}^\mathrm{UV}$ is overestimated
approximately in $10^6 (1-\zeta)$ times due to the screening effect.

This idea concerning the collisional heating of the jet clouds is
observationally supported: $\mathrm{H_\alpha}$ emission of the jets is
anisotropic, the maximum of its directional pattern is directed
to side of jet movement (\cite{Panferov et al. 1997}).
This implies that some screening effect exists for the
impinging protons. It is possible if a magnetic field
tangles trajectories of the protons in a cloud and the stopping column
density $N_\mathrm{s}$ becomes smaller than that given by Eq.(\ref{ns}).
However, the directional pattern can not be explained straightforwardly
as a result of a head-on collision of the clouds with the wind,
because the maximum is inclined to the jet axis.
The inclination of the head maximum is towards the
direction of the precessional motion and is about $40^\circ$ 
to the jet axis. This rather shows that a maximum outcome of
the emission line radiation is from jet sides moving in the wind.

As a result of collision with the wind a cloud decelerates at a rate:
\begin{equation}
\label{dec}
\frac{\Delta v}{v} \approx \frac{\Delta M}{M}= \int \mathrm{d}t
 \frac{\sigma_\mathrm{c} n_\mathrm{w}\, v_\mathrm{j}}{n\, l}=
\int_{R_\mathrm{m}}^{R_\mathrm{j}} \mathrm{d}r \frac{n_\mathrm{w}}
{N_\mathrm{s}} \approx 2\cdot 10^{-2}.
\end{equation}
Here we used $R_\mathrm{m}$ as the lower limit of the integration
because evolution of the jet at distances $< R_\mathrm{m}$ is not well
established. The deceleration (\ref{dec}) agrees well enough
with the observed value of the deceleration $\le 10^{-2}$
(\cite{Kopylov et al. 1987}). Jets sweep out wind gas with a
pattern speed of $\dot \phi\, r$. Then the rate of the heating of 
the whole of optical jet provided by fast protons is:
\begin{eqnarray}
\label{hj}
H = \int_{R_\mathrm{in}}^{R_\mathrm{j}} \mathrm{d}r\,
\frac{N_\mathrm{f}}{N_\mathrm{s}}
\epsilon_\mathrm{p} n_\mathrm{w} (\dot \phi\, r)(
\theta_\mathrm{j} r) \approx 2.2\cdot 10^{37}\ {\rm erg\,s^{-1}}.
\end{eqnarray}
This integral is calculated numerically and 
dependence of $N_\mathrm{f}$ on a geometry of the precessing conical jet
is taken into account. The calculated value of the whole heat $H$
agrees with the observable radiation output 
of the jets in H$_\alpha$ line $L_\mathrm{H_\alpha}\approx 10^{36}$
erg\,s$^{-1}$ (\cite{Panferov et al. 1997}): a part $< 0.1$ of absorbed
energy is emitted in H$_\alpha$.

The small clouds in question would be overheated and their H$_\alpha$
radiation would have ceased unless the clouds are dense enough
to establish a balance between radiative losses and the heating by
the fast protons. In the temperature region near $10^4$ K, which is the
threshold ionization temperature of hydrogen, the efficiency of the 
radiative cooling $\lambda(T)$ strongly depends on temperature 
and changes by 3 orders of magnitude: $\lambda= 10^{-25}$ -- $10^{-22}\,
\mathrm{erg\, cm^3\, s^{-1}}$
(\cite{Kaplan Pikelner 1979}).
From this and Eq.(\ref{dyn}) one finds limits of possible values of the 
radiative energy loss of the clouds, corresponding to 
the end and the beginning of the jet:
\begin{equation}
\label{clim}
3\cdot 10^{-3}\ {\rm erg\,cm^{-3}\,s^{-1}} \le \lambda n^2 \le 9\cdot 10^4\
{\rm erg\,cm^{-3}\,s^{-1}}.
\end{equation}
Since limits on the heating rate $H_c$ from Eq.(\ref{hc}) correspond to:
\begin{equation}
\label{hlim}
9\cdot 10^{-2}\ {\rm erg\,cm^{-3}\,s^{-1}} \le H_\mathrm{c}/l^3
\le 3\cdot 10^3\ {\rm erg\,cm^{-3}\,s^{-1}},
\end{equation}
for the jet end and beginning respectively,
the heating rate never can exceed the cooling capability, given by
inequality (\ref{clim}), and the thermal balance of the clouds can be 
maintained near a temperature of $10^4$ K throughout the whole of the
optical jets.

\section{Evolution and confinement of clouds}
The lifetime of the clouds in the jets, about 4 days, is much
longer than the cloud sound-crossing time. Therefore, the clouds must be 
in pressure equilibrium with the ambient medium and should not be exposed
to destructive processes. The pressure of the ionizing radiation 
$p_\mathrm{r}=H_\mathrm{r}^\mathrm{UV}/l^2 c\approx 9\,r_{14}^{-2}$
dyn is much smaller
than the gas pressure in a cloud $p=2\,n\,k\,T\approx 10^3\,
r_{14}^{-2}$ dyn, where we use a temperature of $T=20000$ K
(\cite{Panferov 
Fabrika 1997}). Therefore, radiation is not important for
dynamical balance of a cloud. The pressure equilibrium of 
clouds with surroundings imposes constraints on density
$n_\mathrm{h}=n\,f \sim 10^7 \ {\rm cm}^{-3}$ and on temperature
$T_\mathrm{h}=T\, n/n_\mathrm{h} \sim 10^{10}$ K of the surrounding gas at
a distance $R_\mathrm{m}$. In this hot medium the clouds will evaporate
with a timescale of:
\begin{equation}
\label{tev}
t_\mathrm{ev}=\frac{m}{\dot m} = \frac{\rho \,l}{6\,\rho_\mathrm{h}
\, c_\mathrm{h}\, F(\sigma_0)}
\ll 10^4\ {\rm s},
\end{equation}
where $m$ is the mass of the cloud, $c_\mathrm{h}$ is sound velocity in
the hot medium, $F(\sigma_0)$ is some function (given by equation
(62) in \cite{Cowie McKee 1977}), which is $\gg 1$ for our case.
The evaporation of the clouds with so short timescale contradicts
to the lifetime of the clouds. Therefore the clouds evaporation 
must be suppressed. 
On the other hand, the ram pressure of the wind is
important in pressure balance of the clouds:
$ p_\mathrm{w}/p=\rho_\mathrm{w}\, v_\mathrm{j}^2/2\, n\, k\, T \approx 17$.
Ram pressure of the wind gas seems to be strongest and must govern
the clouds pressure. But again, gas-dynamical instabilities
are capable of disrupting the clouds on a timescale given by Eq.(\ref{tcr}).
This is the puzzle of SS\,433 jets: how can thermal gas survive there?

\begin{figure}
\resizebox{\hsize}{!}{\includegraphics{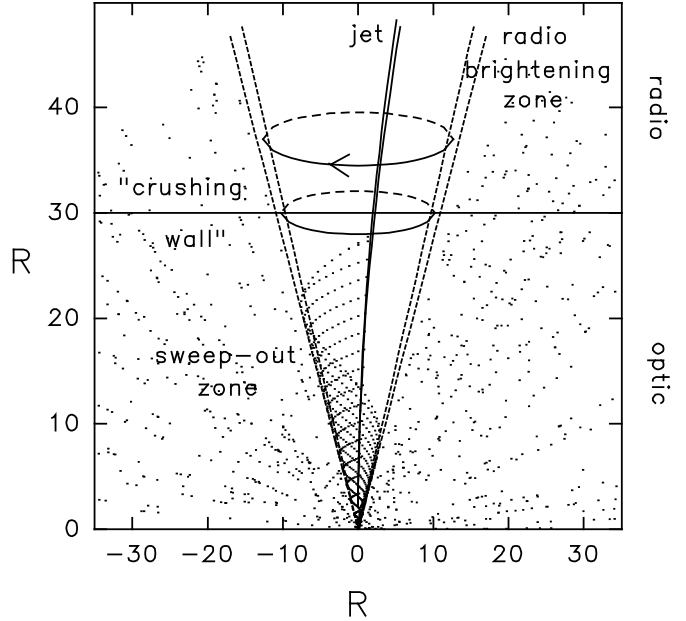}}
\caption{Sketch of precessing jet of SS\,433. Horizontal axis
lies in the accretion disk plane. Both axes are in 
units of $10^{14}$ cm. Scales are exact. The wind of the disk
fills all space except the jet precession surface after distance
$R_\mathrm{sw} \approx 3\cdot 10^{15}$ cm, that is the distance 
of ``crushing wall''. And in bounds of the sweep-out zone
the wind fills the jet precession surface only along a helical pattern moving 
with the precession period. The expansion of the optically
radiating clouds occurs beyond ``crushing wall'' between
two rings.}
\label{wall}
\end{figure}

The optical jets end at a distance of about $3\cdot 10^{15}$ cm.
Further, at a distance $3.7\cdot 10^{15}$ cm, the jets brighten in radio
which is a consequence of jet expansion (\cite{Vermeulen et al. 1987}).
It seems that the expansion is connected with the termination
of the optical jet.
As long as the jets precess and move through the dense wind of
the accretion disk they sweep out  the wind gas to the
surface of the precession cone. After the passage of the jets the wind
fills the jet channels and during the precessional period a region  
of length $R_\mathrm{sw}=P_\mathrm{pr} v_\mathrm{w}
\approx 3\cdot 10^{15}$ cm can be refilled.
Consequently, at distances larger than the sweep-out distance the
lobe along the precession cone surface is empty (see Fig.~\ref{wall}).
Thermal diffusion of wind gas is not capable of filling the lobe:
$c_\mathrm{h} P_\mathrm{pr}
\ll \theta_\mathrm{j} r$ at $r>R_\mathrm{sw}$.
The dependences (\ref{dyn}) of the clouds parameters on a distance
and the clouds heating balance have not peculiarities in point
$r=R_\mathrm{sw}$. But external to the clouds conditions
abruptly change there. Possibly, this forces the confinement of
the clouds to switch off and results in the expansion of the jet. 

From laws in Eqs.(\ref{dyn}) we derive the approximate size of the cloud
$l\approx 4\cdot 10^8$ cm and the filling factor $f\approx 5\cdot 10^{-7}$
at the end of the optical jets.
With these parameters the clouds start to 
expand freely and fill the whole volume of the jets in time:
\begin{equation}
\label{tex}
t_\mathrm{ex} \sim \frac{l}{c_\mathrm{s} f^{1/3}}
\approx 0^\mathrm{d}.5,
\end{equation}
where $c_\mathrm{s}$ is sound velocity in the cloud.
The estimated time of expansion corresponds well to the 
time interval required for the jet to move between the end of 
the optical part and the radio brightening zone.
The last could develop due to the increase of the rate 
of generation of relativistic particles, 
when the clouds expand and turbulence is enhanced.

The {\em coincidence} of lengths of the optical jets and the sweep-out zone
indicates that ram pressure of the wind gas causes
confinement of the clouds. The ram pressure switch off is equivalent to
a ``hard wall'' which crushes the clouds. However, the ram pressure
by itself is not able to prevent clouds from stripping off the gas and
thermal evaporation. On the contrary, it seems from Eq.(\ref{tcr}) that
the higher the ram pressure the shorter the lifetime of a cloud. In reality,
ionized clouds and their
surroundings harbour magnetic fields, which should be important
for stability of the clouds. The magnetic field inferred from radio
observations is $B\approx 0.08$ G at the distance of radio brightening
(\cite{Vermeulen et al. 1987}). Magnetic flux constancy implies that
$B$ scales as $r^{-1}$ and we assume magnetic field in the
jet is:
\begin{equation}
\label{mf}
B = 3\, r_{14}^{-1}\ {\rm G}.
\end{equation}
This field could not confine the clouds having internal pressure
$\sim 10^3 r_{14}^{-2}$ dyn. Jones et al. (\cite{Jones et al. 1996})
showed in their numerical simulations that a ``magnetic shield''
forms around a cloud moving through a medium. The field lines of the
medium are caught by a cloud, stretched and the magnetic field is amplified.
The ``shield'' has magnetic pressure comparable  to the ram pressure and
is able to quench gas-dynamical instabilities and 
prevent evaporation of a cloud.
Energy of the ``magnetic shield'' is converted from the kinetic energy
of the cloud.
So, the magnetic field needed for the confinement of the clouds in the
jets of SS\,433 can be generated by the motion of the clouds
through the wind of the accretion disk. This mechanism of the confinement
naturally explains the termination of the optical jet beyond the
``crushing wall'' at a distance $R_\mathrm{sw}$, where collisional 
interaction of the clouds with the wind ceases.

\section{Conclusions}
We have considered consequences of collisional
interaction of clouds in the jets of SS\,433
with the wind of the accretion disk. This interaction
appears to be a determining factor of  the clouds state and evolution.
This is emphasized by the fact that the clouds exist only
in boundaries of the sweep-out zone over which the jet sweeps out
the wind. We propose that in the 
sweep-out zone the clouds are prevented from destruction 
essentially by a magnetic field which is amplified due to a collisional
interaction with the wind.
The rapid expansion of the clouds after their exit from the sweep-out
zone naturally explains the disappearance of the jet hydrogen line
emission and the brightening of the jets at radio wavelengths.

The collisional interaction of the jets of SS\,433 with its surroundings
is possible for two reasons: the powerful wind from the supercritical
accretion disk and the precession with cone opening being much 
larger than the jet opening angle. These factors are 
necessary for the existence of the
cold clouds radiating the hydrogen lines. Their combination
may be the clue to the uniqueness of the optical relativistic jets of
SS433.

\begin{acknowledgements}
\noindent
We thank D. Baiko for helpful comments.
This research was supported by Polish KBN
Research Grant 2PO3D-021-12 and by
Grant 96-02-16396 of Russian Fond of Basement Researches.
\end{acknowledgements}

\end{document}